# Effect of superthermal electrons on the Quantum electron acoustic double layers in dense astrophysical plasmas

Aakanksha Singh and Punit Kumar[*]

*Department of Physics, University of Lucknow, Lucknow-226007, India*

kumar_punit@lkouniv.ac.in

**ABSTRACT**

Electron-acoustic double layers (EADLs) have been investigated in four component unmagnetized dense quantum plasmas consisting of stationary background ions and two electron populations, 'cold' and 'hot' with the superthermal kappa-distributed electrons. Using the quantum hydrodynamic (QHD) model and the reductive perturbation technique, a generalized Korteweg–de Vries (KdV) equation was derived, and stationary analytical solutions are obtained. The analysis revealed that superthermal electrons substantially influence the amplitude, width, and polarity of EADLs. Numerical results indicated that decreasing the spectral index $\kappa$ or increasing the relative density of $\kappa$-electrons to hot electrons intensifies nonlinear effects, producing stronger compressive and rarefactive structures. It is also found that $\kappa$ plays a more dominant role than density ratio in controlling EADL properties in dense astrophysical environments.

# 1. Introduction

Over the last few years, there has been growing interest in the study of quantum plasmas [1–9] due to their diverse applications in both natural environments and laboratory settings. Quantum plasmas are relevant in a variety of physical systems, including planetary interiors, compact astrophysical bodies [10], and ultracold plasmas [11], as well as in advanced technological environments such as semiconductors, micro-electromechanical systems [12], high-intensity laser-plasma interactions, quantum x-ray free-electron lasers [13], nanoscale electronic devices, and quantum computing platforms [14]. Astrophysical quantum plasmas are typically found in the interiors of dense astrophysical objects like white dwarfs, neutron stars and magnetars [15–17] where the plasma is extremely dense and strongly degenerate [18,19]. At such high densities, the Fermi temperature is typically much greater than the system's thermal temperature, and the de Broglie wavelength of charge carriers becomes comparable to their mean interparticle distance [20]. Under these conditions, the plasma behaves like degenerate Fermi gas, with quantum mechanical effects playing a crucial role in determining the dynamics of the charged particles [21–23]. The dependence of de Broglie wavelength upon mass of the particle and thermal energy shows that quantum effects are more significant for electrons than ions due to smaller mass of electron [24].

Electron acoustic waves (EAWs) are high-frequency dispersive plasma modes (relative to the ion plasma frequency), characterized by the oscillation of a small fraction of inertial cold electrons against a predominant population of inertialess hot electrons [25–27]. These waves have been experimentally observed in laser-produced plasmas and are also relevant in dense astrophysical environments, where the existence of two distinct electron populations is well established [28, 29]. In last few years, both linear and nonlinear aspects of

EAW propagation have been extensively investigated in unmagnetized and magnetized quantum plasmas, considering both planar and nonplanar geometries [30–33].

Over the past decade, considerable attention has been directed toward the study of linear and nonlinear electron acoustic waves [34–36], as well as other wave phenomena such as ion acoustic [37], dust acoustic [38,39], shock waves [40, 41], and Alfvénic waves [42], within the rapidly developing field of quantum plasmas. These investigations have been conducted primarily within the framework of quantum hydrodynamic and quantum magnetohydrodynamic models [43–45], which represent quantum extensions of the conventional classical fluid models. On another side a number of studies have also been devoted to the investigation of double layers (DLs) in quantum plasmas [46-59]. DLs are nonlinear electrostatic structures characterized by a monotonic variation in the electrostatic potential and associated plasma parameters, transitioning from one extreme value to another. Such potential transitions, confined to narrow spatial regions, are capable of accelerating and energizing charged particles. In this context, only a limited number of studies have investigated electron acoustic double layers (EADLs) in dense astrophysical plasmas using the quantum hydrodynamic (QHD) and quantum magnetohydrodynamic model (QMHD) [24, 60,61].

Theoretical investigations of electron acoustic double layers (EADLs) in quantum plasmas have predominantly employed models based on the standard Fermi distribution function for electrons. However, in quantum astrophysical plasmas, characterized by extremely high particle densities and Fermi temperatures that greatly exceed the ambient thermal temperature, electron distribution functions may deviate from the conventional Fermi-Dirac or Bose-Einstein forms. In such environments, the presence of superthermal particle populations can be more accurately described by the quantum analogue of the Kappa distribution, known as the Kappa-Fermi distribution. This distribution incorporates a power-law tail and accounts for the extended high-energy gap typical of superthermal populations [62]. All theoretical investigations of EADLs have so far been conducted using the classical Kappa distribution function, mainly within the framework of space and astrophysical plasmas [63-73]. However, the investigation of EADLs in quantum plasmas employing the quantum Kappa distribution has not yet been undertaken. Consequently, the use of the quantum Kappa distribution in the analysis of EADLs, especially in the context of dense astrophysical plasmas, constitutes a novel and original aspect of the present study.

In the present work, we investigate the nonlinear behaviour of electron acoustic double layers (EADLs) in a collisionless quantum plasma composed of two electron populations with distinct temperatures namely, cold and hot electrons in the presence of stationary ions. The system incorporates a population of superthermal electrons characterized by a modified Kappa-Fermi distribution. This generalized distribution has been modified to account for the contribution of electrostatic energy, enabling the derivation of an expression for the number density of Kappa-distributed electrons. The theoretical framework is developed using the quantum hydrodynamic (QHD) model, which effectively captures the dynamics of quantum plasma constituents and facilitates the analysis of nonlinear structures and collective excitations [74]. The QHD model is particularly well-suited for astrophysical contexts, where quantum effects are essential for accurately describing the behaviour of degenerate electrons and high-density ion dynamics in compact objects such as white dwarfs and neutron stars [75,76], as well as in stellar evolution scenarios [77]. To describe the small-amplitude double layers in an unmagnetized plasma, absent of dissipation and geometrical complexities, we employ a generalized Korteweg–de Vries (KdV) equation. This equation provides a simplified yet robust framework for analyzing the formation, propagation, and dependence of EADLs on plasma parameters. Moreover, we examine the influence of key parameters including the equilibrium density ratio of Kappa electrons to hot electrons

$\left(\Delta = n_{\kappa e0}/n_{he0}\right)$ and the Kappa index (κ) on the characteristics of the Sagdeev potential as well as EADL solutions.

The remaining part of this paper is organized as follows; In Section. 2, we consider model equations. In Section. 3 we present the nonlinear analysis and DLs solution. Finally, Section. 4 is devoted to summary and discussion.

## 2. Governing set of equations

Consider the propagation of EAWs in four component dense quantum plasma consisting of a population of inertial cold electrons, inertialess hot electrons, Kappa distributed electrons and stationary ions forming the neutralizing charge background. These fours plasma species are henceforth denoted by c, h, $\kappa$ and i, respectively. Thus the basic set of equations governing the dynamics of EAWs in collisionless quantum plasma are [30,31],

$$\frac{\partial n_\alpha}{\partial t} + \frac{\partial}{\partial x}\left(n_\alpha \vec{v}_\alpha\right) = 0, \tag{1}$$

$$\frac{\partial \vec{v}_\alpha}{\partial t} + \vec{v}_\alpha \frac{\partial \vec{v}_\alpha}{\partial x} = \frac{e_\alpha}{m_\alpha}\frac{\partial \phi}{\partial x} - \frac{1}{m_\alpha n_\alpha}\frac{\partial P_{F\alpha}}{\partial x} + \frac{\hbar^2}{2m_\alpha^2}\frac{\partial}{\partial x}\left(\frac{1}{\sqrt{n_\alpha}}\frac{\partial^2}{\partial x^2}\sqrt{n_\alpha}\right), \tag{2}$$

and

$$\frac{\partial^2 \phi}{\partial x^2} = \frac{e}{\varepsilon_0}\left(n_{ce} + n_{he} + n_{ke} - Z_i n_{0i}\right). \tag{3}$$

Equations (2) and (3) represent the continuity and momentum equations for the $\alpha$ - species ($\alpha = c, h$) in the plasma, respectively. In Equation (2), the second term on the left-hand side denotes the convective derivative of the fluid velocity, while the first term on the right-hand side corresponds to the electrostatic force, expressed as $E = -\nabla \phi$, where $\phi$ is the electrostatic potential. The second term on the right-hand side accounts for the force due to Fermi pressure, and the third term represents the quantum Bohm potential, which arises from quantum corrections associated with density fluctuations. These corrections, originating from the wave-like nature of the charge carriers, are collectively referred to as quantum diffraction effects. Here, ℏ denotes the reduced Planck's constant. The densities of Kappa-distributed hot electrons, inertial cold electrons, inertialess hot electrons, and stationary ions are coupled through the Poisson equation (3).

In equilibrium, the plasma holds quasi-neutrality condition, $n_{ce0} + n_{he0} + n_{\kappa e0} = z_i n_{i0}$ where $n_{ce0}$, $n_{he0}$ and $n_{\kappa e0}$ are equilibrium densities of cold, hot and kappa electrons. Here $n_{i0}$ is the background ion. In EAWs, the cold electrons provide the inertia and hot electrons the restoring force, respectively. The phase speed of the EAW lies in the range $v_{Fce} \ll \omega/k \ll v_{Fhe}$, where $v_{Fce}$ and $v_{Fhe}$ are the Fermi velocities of cold and hot electrons, respectively. EAW propagates on cold electron dynamic scale with $n_{ce0} \ll n_{he0}$ and the plasma frequency due to hot and cold electrons is defined as $\omega_{pe\alpha} = (n_{\alpha e0} e^2 / m_e)^{1/2}$ therefore, the condition $\omega_{pce} \ll \omega_{phe}$ holds for EAWs in quantum plasmas. The electron acoustic speed is defined as $c_{ea} = (2k_B T_{Fhe} \delta / m_e)^{1/2}$ where, $\delta = n_{ce0}/n_{he0} < 1$ and $\lambda_{Fhe} = (2K_B T_{Fhe}/n_{he0} e^2)^{1/2} = v_{Fhe}/\omega_{phe}$ is the Fermi wavelength due to hot electrons in quantum plasma.

For hot electrons, the equation of state is described by the one dimensional quantum Fermi-gas model which is given as, $P_{Fhe} = m_{he} v_{Fhe}^2 n_{he}^3 / 3n_{he0}^2$ [74] and derived under the assumption of a zero-temperature Fermi-Dirac distribution for electrons. Additionally, the dimensionless quantum diffraction parameter is introduced, $H = \hbar \omega_{pe\alpha} / m_{e\alpha} v_{Fe\alpha}^2$ where,

$v_{F\alpha} = \sqrt{2k_B T_{F\alpha}/m_{e\alpha}}$ is Fermi velocity. In high-density plasmas, the stabilizing contributions arise from both the Fermi pressure term proportional to $v_{F\alpha}^2$ and the Bohm pressure term proportional $H^2$. Consequently, from Eq. (2), the momentum equations governing the cold electron dynamics and the inertialess hot electrons can be formulated accordingly,

$$\frac{\partial \vec{v}_c}{\partial t} + \vec{v}_c \frac{\partial \vec{v}_c}{\partial x} = \frac{e}{m_e} \frac{\partial \phi}{\partial x} + \frac{H_{ce}^2 v_{Fce}^4}{2\omega_{Pce}^2} \frac{\partial}{\partial x}\left(\frac{1}{\sqrt{n_{ce}}} \frac{\partial^2}{\partial x^2} \sqrt{n_{ce}}\right),\tag{4}$$

and

$$0 = \frac{e}{m_e} \frac{\partial \phi}{\partial x} - \frac{v_{Fhe}^2}{n_{he0}^2} n_{he} \frac{\partial n_{he}}{\partial x} + \frac{H_{he}^2 v_{Fhe}^4}{2\omega_{Phe}^2} \frac{\partial}{\partial x}\left(\frac{1}{\sqrt{n_{he}}} \frac{\partial^2}{\partial x^2} \sqrt{n_{he}}\right).\tag{5}$$

Since the conditions $n_{ce0} \ll n_{he0}$ and $T_{Fce} \ll T_{Fhe}$ must hold for the electrostatic wave (EAW) in quantum plasmas, it follows that, the Fermi pressure contributed by cold electrons have been neglected relative to the pressure from hot electrons in the model (as in eq. (4)). Additionally, since the phase speed of the EAW lies within a certain range $v_{Fce} \ll \omega/k \ll v_{Fhe}$, the inertia of the hot electrons has been considered negligible in this model (as in eq. (5)).

We consider a significant population of hot electrons characterized by a high Fermi temperature, $T_{Fe} \approx 10^7 - 10^8 K$ and correspondingly elevated thermal energy. These electrons are described by the Kappa-Fermi distribution (also referred to as the $\kappa$-Fermi distribution) and are commonly known as Kappa electrons, or superthermal electrons. The Kappa parameter characterized by $(\kappa < 1)$, in quantum plasmas, quantifies the extent to which these electrons deviate from the Fermi-Dirac distribution, accounting for the presence of a high-energy tail in their distribution function. The generalized form of three dimensional Kappa-Fermi distribution in the presence of electrostatic potential is [details in appendix],

$$f_{\kappa e}(\phi) = \frac{n_{eo}\left\{2-\mu\beta\left(1+\frac{s}{\kappa}\right)\right\}^{\left(\frac{1}{\kappa+s}-\frac{1}{2}\right)} \Gamma\left(\frac{1}{\kappa+s}+1\right)}{\pi^{3/2}\left[\frac{\left(1+\frac{s}{\kappa}\right)\beta}{2m_e}\right]^{-3/2} \Gamma\left(\frac{1}{\kappa+s}-\frac{1}{2}\right)} \left[1+\left\{1+\frac{\beta}{\kappa}\left(\frac{p_e^2}{2m_e}+q_e\phi-\mu\right)\right\}^{\kappa+s}\right]^{-\left(1+\frac{1}{\kappa+s}\right)}.$$

(6)

where, $\mu = \frac{A_\kappa^{2/3} \hbar^2 k_F^2}{2m}$ represents the chemical potential for Olbert-Fermi gas [62], $n_{\kappa e0}$ is the equilibrium density of Kappa distributed electrons, $A_\kappa = 2^{1/(\kappa+s)}$ is modification term which depends upon Kappa parameter and $k_F = \left(3\pi^2 n_{\kappa e0}\right)^{1/3}$ is the Fermi wavenumber. This distribution preserves the quantum characteristics of the electron gas, which become especially relevant at low temperatures. A fixed positive constant, s > 0 is introduced to account for thermodynamic constraints. The value of this constant depends on the nature of the gas, which for an ideal non-relativistic gas, s = 5/2 and for an ideal relativistic gas, s = 4, as established by thermodynamic principles [78,79]. In the limit of large κ ($\kappa \gg 1$), the deviation from the standard Fermi-Dirac distribution becomes negligible i.e., the distribution asymptotically approaches the ordinary Fermi-Dirac form. By integrating the κ-Fermi distribution over momentum space, one obtains the number density of the degenerate electron population, which explicitly depends on the electrostatic potential. Consequently, the number density of κ-distributed (superthermal) electrons $\left(q_e = -e\right)$ can be expressed as,

$$n_{\kappa e}(\phi) = \frac{n_{\kappa e 0}}{4\pi^3 h^3 A_\kappa} \left[ \frac{1 - \frac{(e\phi + \mu)}{2}\left(1 + \frac{s}{\kappa}\right)\beta}{1 - \frac{\mu\beta}{2}\left(1 + \frac{s}{\kappa}\right)} \right]^{\frac{1}{2} - \frac{1}{\kappa + s}},$$

(7)

Now substituting the value of density of kappa electrons from eq. (7), the Poission equation (3) becomes

$$\frac{\partial^2 \phi}{\partial x^2} = \frac{e}{\varepsilon_0}\left[ n_{ce} + n_{he} + \alpha_\kappa \left\{ 1 - \left(\frac{\kappa + s - 2}{4\kappa}\right)(e\phi + \mu)\beta \right\} - Z_i n_{i0} \right],$$

(8)

where, $\alpha_\kappa = \dfrac{n_{\kappa e 0}}{4\pi^3 h^3 A_\kappa}\left[1 - \dfrac{\mu\beta}{2}\left(1 + \dfrac{s}{\kappa}\right)\right]^{\frac{1}{\kappa + s} - \frac{1}{2}}.$

Above equation corresponds to Poission's equation by which densities of the constituent particles (Kappa-distributed hot electrons, inertial cold electrons, inertialess hot electrons, and stationary ions) have been coupled. The equations (1), (4), (5) and (8) are together referred as the four set of governing equations describing the dynamics of the plasma system.

## 3. Analysis for Double Layer Structures

In order to derive the nonlinear equation and governing the dynamics of weak electron acoustic double layers, we introduce the stretched variables in space and time as

$$\zeta = \varepsilon(x - \lambda t),$$

$$\tau = \varepsilon^3 t,$$

(9)

and expand the field quantities $n_\alpha$, $v_\alpha$ and $\phi$ about their equilibrium values in the power of $\varepsilon$ as

$$\begin{pmatrix} n_\alpha \\ v_\alpha \\ \phi \end{pmatrix} = \begin{pmatrix} n_{0\alpha} \\ 0 \\ 0 \end{pmatrix} + \varepsilon \begin{pmatrix} n_\alpha^{(1)} \\ v_\alpha^{(1)} \\ \phi^{(1)} \end{pmatrix} + \varepsilon^2 \begin{pmatrix} n_\alpha^{(2)} \\ v_\alpha^{(2)} \\ \phi^{(2)} \end{pmatrix} + \varepsilon^3 \begin{pmatrix} n_\alpha^{(3)} \\ v_\alpha^{(3)} \\ \phi^{(3)} \end{pmatrix}.$$

(10)

where, $\varepsilon$ is the small parameter measuring the amplitude of perturbation as well as strength of nonlinearity. $\lambda$ is the phase speed of the wave. From equation (9), the dependent variables $x$ and $t$ are functions of $\zeta$ and $\tau$. Therefore, equations (1), (4), (5) and (8) can be transformed into stretched co-ordinates $\zeta$ and $\tau$ as,

$$0 = -\varepsilon\lambda \frac{\partial n_\alpha}{\partial \zeta} + \varepsilon^3 \frac{\partial n_\alpha}{\partial \tau} + \varepsilon \frac{\partial}{\partial \zeta}(n_\alpha v_\alpha),$$

(11)

$$-\varepsilon\lambda \frac{\partial v_c}{\partial \zeta} + \varepsilon^3 \frac{\partial v_c}{\partial \tau} + \varepsilon v_c \frac{\partial v_c}{\partial \zeta} = \varepsilon \frac{e}{m_e}\frac{\partial \phi}{\partial \zeta} + \varepsilon^3 \frac{H_{ce}^2 v_{Fce}^4}{2\omega_{Pce}^2}\frac{\partial}{\partial \zeta}\left(\frac{1}{\sqrt{n_{ce}}}\frac{\partial^2}{\partial \zeta^2}\sqrt{n_{ce}}\right),$$

(12)

$$0 = \varepsilon \frac{e}{m_e}\frac{\partial \phi}{\partial \zeta} + \varepsilon^3 \frac{H_{he}^2 v_{Fhe}^4}{2\omega_{Phe}^2}\frac{\partial}{\partial \zeta}\left(\frac{1}{\sqrt{n_{he}}}\frac{\partial^2}{\partial \zeta^2}\sqrt{n_{he}}\right) - \varepsilon \frac{v_{Fhe}^2}{n_{he0}^2}\left(n_{he}\frac{\partial n_{he}}{\partial \zeta}\right),$$

(13)

and

$$\varepsilon^2 \frac{\partial^2 \phi}{\partial \zeta^2} = \frac{e}{\varepsilon_0}\left[n_{ce} + n_{he} - Z_i n_{i0} + \alpha_\kappa\left\{1 - \left(\frac{\kappa + s - 2}{4\kappa}\right)(e\phi + \mu)\beta\right\}\right].$$

(14)

Applying the perturbative expansion of the field quantities in the above transformed set of equations and taking the lowest order terms of $\varepsilon$, we get the following set of equations

$$n_{0\alpha}\frac{\partial v_\alpha^{(1)}}{\partial \zeta} - \lambda \frac{\partial n_\alpha^{(1)}}{\partial \zeta} = 0,$$

(15)

$$\frac{e}{m_e}\frac{\partial \phi^{(1)}}{\partial \zeta} + \lambda \frac{\partial v_c^{(1)}}{\partial \zeta} = 0,$$

(16)

$$\frac{e}{m_e}\frac{\partial \phi^{(1)}}{\partial \zeta} - \frac{v_{Fhe}^2}{n_{he0}}\frac{\partial n_{he}^{(1)}}{\partial \zeta} = 0,$$

(17)

and

$$\frac{e}{\varepsilon_0}\left[n_{ce}^{(1)} + n_{he}^{(1)} - e\alpha_\kappa\beta\left(\frac{\kappa+s-2}{4\kappa}\right)\phi^{(1)}\right] = 0.$$

(18)

Integrating and solving the above set of equations, the following expressions for the perturbed quantities are obtained as,

$$n_c^{(1)} = \frac{n_{ce0}}{\lambda}v_c^{(1)},$$

(19)

$$v_c^{(1)} = -\frac{e}{m_e\lambda}\phi^{(1)},$$

(20)

$$n_c^{(1)} = -\frac{en_{ce0}}{m_e\lambda^2}\phi^{(1)},$$

(21)

$$n_h^{(1)} = \frac{en_{he0}}{m_e v_{fe}^2}\phi^{(1)}.$$

(22)

with the phase speed of the wave given by,

$$\lambda^2 = \frac{en_{ce0}}{m_e\left(\frac{en_{he0}}{m_e v_{Fhe}^2} - e\alpha_\kappa\beta\left(\frac{\kappa+s-2}{4\kappa}\right)\right)}.$$

(23)

Now, equating the next higher order terms of $\varepsilon$, we get

$$-\lambda\frac{\partial n_\alpha^{(2)}}{\partial\zeta} + n_{o\alpha}\frac{\partial v_\alpha^{(2)}}{\partial\zeta} + \frac{\partial}{\partial\zeta}\left(n_\alpha^{(1)}v_\alpha^{(1)}\right) = 0,$$

(24)

$$-\lambda\frac{\partial v_c^{(2)}}{\partial\zeta} + v_c^{(1)}\frac{\partial v_c^{(1)}}{\partial\zeta} = \frac{e}{m_e}\frac{\partial\phi^{(2)}}{\partial\zeta},$$

(25)

$$\frac{e}{m_e}\frac{\partial\phi^{(2)}}{\partial\zeta} - \frac{v_{Fhe}^2}{n_{he0}^2}\frac{\partial n_{he}^{(2)}}{\partial\zeta} - \frac{v_{Fhe}^2}{n_{he0}^2}\left(n_{he}^{(1)}\frac{\partial n_{he}^{(1)}}{\partial\zeta}\right) = 0,$$

(26)

and

$$0 = \frac{e}{\varepsilon_0}\left[n_{ce}^{(2)} + n_{he}^{(2)} - e\alpha_\kappa\left(\frac{\kappa+s-2}{4\kappa}\right)\beta\phi^{(2)}\right].\tag{27}$$

Integrating these equations and then solving them, we obtain following set of equations,

$$v_c^{(2)} = \frac{1}{2\lambda}\left(v_c^{(1)}\right)^2 - \frac{e}{m_e\lambda}\phi^{(2)},\tag{28}$$

$$n_c^{(2)} = \frac{n_{ce0}}{\lambda}v_c^{(2)} - \frac{n_c^{(1)}v_c^{(1)}}{\lambda},\tag{29}$$

$$n_h^{(2)} = \frac{en_{he0}}{m_e v_{fe}^2}\phi^{(2)} + \frac{1}{2n_{he0}}\left(n_h^{(1)}\right)^2,\tag{30}$$

and

$$\left[n_{ce}^{(2)} + n_{he}^{(2)} - e\alpha_\kappa\left(\frac{\kappa+s-2}{4\kappa}\right)\beta\phi^{(2)}\right] = 0,\tag{31}$$

Substituting equations (20)-(23) into equations (28)-(31), we obtain

$$\chi\left(\phi^{(1)}\right)^2 = 0\tag{32}$$

where, $\chi = \frac{e^2}{2m_e^2}\left(\frac{n_{he0}}{v_{fe}^4} - \frac{3n_{ce0}}{\lambda^4}\right)$. Since $\phi^{(1)} \neq 0$, $\chi$ should be at least of the first order of $\varepsilon$ and so $\chi\left(\phi^{(1)}\right)^2$ has to be substituted in the next higher order equation for $\phi$. Now equating the next higher order terms of $\varepsilon$ we obtain the following set of equations

$$-\lambda\frac{\partial n_\alpha^{(3)}}{\partial \zeta} + \frac{\partial n_\alpha^{(1)}}{\partial \tau} + \frac{\partial}{\partial \zeta}\left(n_\alpha^{(1)}v_\alpha^{(2)}\right) + \frac{\partial}{\partial \zeta}\left(n_\alpha^{(2)}v_\alpha^{(1)}\right) + n_{o\alpha}\frac{\partial v_\alpha^{(3)}}{\partial \zeta} = 0,\tag{33}$$

$$-\lambda\frac{\partial v_c^{(3)}}{\partial \zeta} + \frac{\partial v_c^{(1)}}{\partial \tau} + v_c^{(1)}\frac{\partial v_c^{(2)}}{\partial \zeta} + v_c^{(2)}\frac{\partial v_c^{(1)}}{\partial \zeta} = \frac{e}{m_e}\frac{\partial \phi^{(3)}}{\partial \zeta} + \frac{H_{ce}^2 v_{Fce}^4}{4\omega_{Pce}^2 n_{ce0}}\frac{\partial^3 n_{ce}^{(1)}}{\partial \zeta^3},\tag{34}$$

$$\frac{e}{m_e}\frac{\partial \phi^{(3)}}{\partial \zeta} - \frac{v_{Fhe}^2}{n_{0h}}\frac{\partial n_{he}^{(3)}}{\partial \zeta} - \frac{v_{Fhe}^2}{n_{0h}^2}\left(n_{he}^{(1)}\frac{\partial n_{he}^{(2)}}{\partial \zeta}\right) - \frac{v_{Fhe}^2}{n_{0h}^2}\left(n_{he}^{(2)}\frac{\partial n_{he}^{(1)}}{\partial \zeta}\right) + \frac{H_{he}^2 v_{Fhe}^4}{4\omega_{Phe}^2 n_{he0}}\frac{\partial^3 n_{he}^{(1)}}{\partial \zeta^3} = 0,$$
(35)

and

$$\frac{\partial^2 \phi^{(1)}}{\partial \zeta^2} = \frac{e}{\varepsilon_0}\left[n_{ce}^{(3)} + n_{he}^{(3)} - e\alpha_\kappa\left(\frac{\kappa+s-2}{4\kappa}\right)\beta\phi^{(3)} + \chi\left(\phi^{(1)}\right)^2\right].$$
(36)

Using equations (20)-(23) and (28)-(30) in the above set of equations, we finally obtain the required nonlinear KdV equation governing the dynamics of electron acoustic DLs as,

$$\frac{\partial \phi^{(1)}}{\partial \tau} + C_1 \phi^{(1)}\frac{\partial \phi^{(1)}}{\partial \zeta} + C_2\left(\phi^{(1)}\right)^2\frac{\partial \phi^{(1)}}{\partial \zeta} + C_3\frac{\partial^3 \phi^{(1)}}{\partial \zeta^3} = 0$$
(37)

where, $C_1 = \dfrac{e\lambda}{2m_e}\left[\left(\dfrac{\sqrt{3}}{\lambda}\right)^2 - \left(\dfrac{\omega_{phe}\lambda}{\omega_{pce}v_{Fhe}^2}\right)^2\right],$ (38)

$$C_2 = \frac{e^2\lambda}{2m_e^2}\left[\left(\frac{\sqrt{15/2}}{\lambda^2}\right)^2 - \left(\frac{\omega_{phe}\lambda}{\omega_{pce}v_{Fhe}^3}\right)^2\right],$$
(39)

and

$$C_3 = \frac{\lambda}{2}\left[\left(\frac{\lambda}{\omega_{pce}}\right)^2 - \frac{H_{ce}^2 v_{Fce}^4}{4\lambda^2 \omega_{Pce}^2} - \frac{H_{he}^2 \lambda^2}{4\omega_{Pce}^2}\right].$$
(40)

The steady-state solution of equation (37) is obtained by transforming the independent variables $\zeta$ and $\tau$ to $\eta = \zeta - U\tau$ where, $U$ is a normalised constant speed of electron acoustic wave frame. Applying the boundary condition that as $\eta \to \pm\infty$, $\phi \to 0$ and $d\phi/d\eta \to 0$, eq. (37) yields the 'energy integral'

$$\frac{1}{2}\left(\frac{d\phi}{d\eta}\right)^2 + V(\phi) = 0 \tag{41}$$

where,

$$V(\phi) = -\frac{U}{2C_3}\phi^2 + \frac{C_1}{6C_3}\phi^3 + \frac{C_2}{12C_3}\phi^4 \tag{42}$$

is the Sagdeev potential. For double layer solution, this Sagdeev potential should satisfy the condition that as $V(\phi) = 0$, $V'(\phi) = 0$ and $V''(\phi) < 0$, at $\phi = 0$ and $\phi = \phi_m$ [60,61]. Applying these boundary conditions to eq. (42), we obtain the expression for maximum amplitude amplitude ($\phi_m$) and constant speed as

$$\phi_m = -\frac{C_1}{C_2}, \text{ and } U = -\frac{C_2}{6}(\phi_m)^2. \tag{43}$$

Putting these values of $C_1$ and $U$, $V(\phi)$ reduces to

$$V(\phi) = \frac{C_2}{12C_3}\phi^2(\phi_m - \phi)^2. \tag{44}$$

Now integrating eq. (41) with (44), the steady state double layer solution of eq. (37) can be written as

$$\phi = \frac{\phi_m}{2}\left[1 - \tanh\left(\sqrt{-\frac{C_2}{24C_3}}\phi_m\eta\right)\right]. \tag{45}$$

Equation (45) represents a DL provided $(-C_2/C_3 < 0)$. Also the nature of the DL, i.e., whether it is compressive $(\phi > 0)$ or rarefactive $(\phi < 0)$ will be determined by the conditions $(C_1/C_2 < 0)$ or $(C_1/C_2 > 0)$, respectively. The width of the DL is

$$d = \frac{\sqrt{-24C_3/C_2}}{\phi_m} \tag{46}$$

Equations (43)-(46), constitute the main results of our quantum electron acoustic DLs.

In the numerical analysis to follow, the parameters are chosen for dense astrophysical plasmas like white dwarfs and neutron stars, having values in range of $T_{Fe} \approx 10^6 - 10^8 K$, $10^{28} \leq n_{e0} \leq 10^{32} m^{-3}$ so that $0.24 \leq H \leq 1.10$ [80,81] and the value of kappa parameter $\kappa$ for such type of dense astrophysical objects ranges from 0.1 to 0.5. In quantum plasma both the compressive and rarefactive electron acoustic DLs can exist if the quantum effects are taken into account in the electric system, even without retaining the hot electron inertia and external electron beam [60]. Value of the cold to hot electron ratio $(\delta)$, for the compressive and rarefactive double layers in dense astrophysical quantum plasma have been taken as $\delta = 1.75$, $\delta = 2.88$ respectively [60].

Fig. 1(a) shows the variation of Sagdeev potental $V(\phi)$ with electrostatic potential $\phi$, corresponding to compressive double layer in dense astrophysical quantum plasma. The three curves represent distinct κ values, where the solid black curve corresponds to a higher κ (nearly Fermi-Dirac distribution), the dashed blue curve to an intermediate κ, and the dotted black curve to a lower κ (nearly Kappa-Fermi distribution). It is evident from the figure that the depth and width of the potential well are strongly dependent on κ. A decrease in κ (enhanced superthermality) leads to a deeper and broader potential well, indicating stronger nonlinearity and allowing double layers of larger amplitude. In low κ plasmas, the high-energy tail of the distribution function contributes additional free energy, strengthening the electrostatic potential and modifying the balance between nonlinearity and dispersion. Conversely, larger κ values (nearly Fermi-Dirac distribution) yield weaker and more compact (less deep and more confined) potential wells, corresponding to weaker nonlinear effects and reduced double-layer amplitudes.

Fig.1(b) shows the variation of $V(\phi)$ with $\phi$, corresponding to compressive double layer, for different values of the initial number density ratio of kappa eletrons to hot eletrons ($\Delta = n_{\kappa e0}/n_{he0}$ ).It is evident from the figure that the profile of the Sagdeev potential is strongly dependent on $\Delta$. For lower values of $\Delta$ (solid curve), the potential well is relatively shallow, indicating weaker nonlinear interactions and reduced amplitude of the double layer structure. In addition, depth and asymmetry of the potential well increase with higher $\Delta$, suggesting that the plasma system supports stronger electrostatic localized structures under these conditions. Physically, this implies that an enhanced density of suprathermal κ-electrons relative to hot electrons plays a crucial role in intensifying the nonlinear electrostatic interactions, thereby facilitating the existence of stable double layers in dense astrophysical quantum plasmas.

Figs. 2(a) and 2(b) shows the variation of Sagdeev potental $V(\phi)$ with electrostatic potential $\phi$, corresponding to rarefactive double layer. The coexistence of potential maxima and minima in both figures confirms the existence of double layer solutions within the plasma system. Fig 2(a) demonstrates that the Sagdeev potential exhibits a pronounced dependence on the kappa index (κ) parameter. For higher values of κ (weaker superthermality), the potential well is relatively shallow, suggesting weaker nonlinear structures. Conversely, κ as decreases (nearly Kappa-Fermi distribution), the potential becomes markedly deeper and more asymmetric, which reflects the enhanced role of high-energy superthermal particles in steepening the nonlinear electrostatic potential structure thereby indicating the emergence of stronger rarefactive double layer characteristics. From fig.2(b), it is observed that the Sagdeev potential profile is highly sensitive to the variation in Δ. As Δ decreases (i.e., relatively lower kappa electrons), the depth of the Sagdeev potential well decreases, and the profile becomes shallower. This trend suggests that reduction in the presence of κ-electrons reduces the nonlinearity by suppressing the steepness of the potential structure, thereby diminishing the strength of the double layer. Physically, this behaviour highlights the role of the electron population ratio in regulating the existence domain and stability of electrostatic double layers. In addition, we also observe that the $\kappa$- parameter have dominant effects on Sagdeev potential $V(\phi)$ for the both type (compressive and rarefactive) of double layers as compare to density ratio ($\Delta$) [comparing Fig.1 and Fig.2].

Fig.3(a) shows the variation of compressive EADL solution $\phi$ with $\eta$, for different values of kappa index ($\kappa$). This profile demonstrates that the compressive EADL becomes more intense and localized as the kappa index $\kappa$ decreases. Superthermal electrons carry excess kinetic energy, which enhances the nonlinear response of the plasma to perturbations. As $\kappa$ decreases, the high-energy tail of the distribution becomes more pronounced, leading to stronger nonlinearities and more energetic DL formation. The peak value of the electrostatic potential rises significantly which implies that stronger superthermal electron populations support larger-amplitude double layers. The structure is more compressive, with potential increasing over a localized region. This is characteristic of compressive DLs, where denser or more energetic populations pile up over a small region. Therefore, lower $\kappa$ values result in larger and steeper double layers, as seen in the figure.

Fig.3(b) shows the variation of compressive EADL solution $\phi$ with $\eta$, for different values of $\Delta$. This figure clearly illustrates that as the density ratio $\Delta$ increases; the amplitude of the potential across the double layer also increases. Hence, compressive EADL solution becomes stronger and steeper with increasing density ratio of kappa electrons to hot electrons. This suggests that superthermal electron populations also play a significant role in the formation and characteristics of compressive double layers in such plasma systems.

Fig.4(a) shows the variation of rarefactive EADL solution for three distinct values of the index $\kappa$. The monotonic transition in $\eta$ from higher to lower values across the localized region confirms the rarefactive nature of the structure, where electron density depletion is dominant. It is evident from the figure that the amplitude and sharpness of the rarefactive double layer are strongly dependent on the value of the spectral index κ. This figure indicates that the steepness and amplitude of the EADL solution increase with decreasing κ, highlighting the crucial role of superthermal (Kappa-Fermi distributed) population, in determining the strength and stability of rarefactive double layers in dense astrophysical plasma environments.

Fig 4(b) illustrates the variation of the rarefactive EADL solution, for different values of the number density ratio $\Delta$, in dense astrophysical quantum plasma. It is evident that the strength and amplitude of the rarefactive double layer are significantly affected by the variation in $\Delta$. For higher values of $\Delta$, the amplitude of the potential increases, producing a more pronounced rarefactive structure. This behaviour indicates that increasing the relative concentration of κ-electrons enhances the nonlinear steepening of the rarefactive double layer. Conversely, at lower values of $\Delta$, the double layer potential becomes weaker and less steep, suggesting that an deceased proportion of κ-electrons tends to suppress the formation of strong rarefactive structures. In addition, we also observe that the *κ*- parameter have dominant effects on on EADL solution for the both type (compressive and rarefactive) of double layers as compare to density ratio ($\Delta$) [comparing Fig.3 and Fig.4].

## 4. Summary and discussion

This work presented a theoretical investigation of quantum electron acoustic double layers (EADLs) in dense astrophysical plasmas containing cold electrons, hot electrons, stationary ions, and a superthermal electron population characterized by a modified Kappa-Fermi distribution. Using the quantum hydrodynamic (QHD) model in combination with the reductive perturbation technique, a generalized Korteweg–de Vries (KdV) equation was derived to describe weak EADLs, and stationary analytical solutions were obtained. The effects of the kappa index (κ) and the relative density of κ-electrons to hot electrons (Δ) were systematically examined to identify their role in shaping the nonlinear double layer structures.

The analysis shows that superthermal electrons play a decisive role in shaping the amplitude, width, and polarity of double layers. Smaller κ values and higher superthermal electron densities strengthen nonlinearities, producing larger-amplitude compressive and rarefactive structures. In particular, it is found that κ plays a more dominant role than Δ in controlling the strength and profile of EADLs across both compressive and rarefactive regimes.

In conclusion, this study highlights that the presence of superthermal electrons, quantified through the modified Kappa-Fermi distribution, significantly alters the amplitude, width, and polarity of electron acoustic double layers in dense astrophysical plasmas. The dominance of κ in controlling nonlinear effects suggests that the degree of superthermality is the primary factor governing the formation and stability of EADLs. These findings provide deeper insight into localized electrostatic structures in environments such as white dwarf interiors and neutron star crusts, where superthermal populations are expected to be prominent. The developed framework bridges classical superthermal plasma models with quantum plasma theory, offering predictive capability for astrophysical observations and plasma simulations. The developed framework can be extended to magnetized configurations, multi-ion species systems, or relativistic regimes, facilitating the way for deeper understanding of quantum plasma dynamics of compact astrophysical objects.

**CRediT authorship contribution statement**

**Aakanksha Singh**: Writing – review & editing, Writing – original draft, Visualization, Validation, Software, Investigation, Methodology, Formal analysis,

Conceptualization; **Punit Kumar**: Supervision, Resources, Project administration, Funding acquisition.

**Declaration of competing interest**

The authors declare that they have no known competing financial interests or personal relationships that could have appeared to influence the work reported in this paper.

**Acknowledgements**

The authors thank SERB- DST, Govt. of India for financial support under MATRICS scheme (grant no. : MTR/2021/000471).

**Appendix**

Here we provide the complete mathematical derivation of eq. (6) of the manuscript. We start with the Kappa-Fermi distribution for degenerate Olbert-Fermi gases [63],

$$f_\kappa(\varepsilon_p) = A\left[1 + \left\{1 + \frac{\beta}{\kappa}(\varepsilon_p - \mu)\right\}^{\kappa+s}\right]^{-\left(1+\frac{1}{\kappa+s}\right)}, \quad \varepsilon_p > \mu \qquad (i)$$

where, A is a normalization constant, $T \equiv \beta^{-1}$ is a constant physical temperature of olbert Fermi gases with momentum $p$ and particle energy $\varepsilon_p = p^2/2m$. $\mu > 0$ represents the chemical potential, applicable to the Fermi-distribution. The Olbert parameter $\kappa$ accounts for deviations from the standard Fermi distribution, reflecting the influence of internal correlations, or additional degrees of freedom. For $\kappa \to \infty$, the Kappa- Fermi distribution function given in eq. (1) reduces to the standard Fermi-Dirac distribution.

Taking normalization of $f_\kappa(\varepsilon_p)$ over momentum space such that $\int f_\kappa(\varepsilon_p) d^3 p = n_{jo}$, we obtain the following Kappa-Fermi distribution function,

$$f_\kappa(\varepsilon_p) = \frac{n_{jo}\left\{2-\mu\beta\left(1+\frac{s}{\kappa}\right)\right\}^{\left(\frac{1}{\kappa+s}-\frac{1}{2}\right)} \Gamma\left(\frac{1}{\kappa+s}+1\right)}{\pi^{3/2}\left[\frac{\left(1+\frac{s}{\kappa}\right)\beta}{2m_j}\right]^{-3/2} \Gamma\left(\frac{1}{\kappa+s}-\frac{1}{2}\right)} \left[1+\left\{1+\frac{\beta}{\kappa}\left(\frac{p^2}{2m_j}-\mu\right)\right\}^{\kappa+s}\right]^{-\left(1+\frac{1}{\kappa+s}\right)},$$

(ii)

where, $n_{jo}$ is the equilibrium number density, $m_j$ is the mass of the fermion-species $j$ (for example, $j = e^-, e^+$ etc).

In case of presence of electrostatic potential ($\phi$), we have to modify the above distribution to include electrostatic energy contribution. Therefore, using the energy conservation relation $\frac{p^2}{2m_j} = \frac{p_j^2}{2m_j} + q_j\phi$, where $q_j\phi$ is the increase in potential energy due to presence of electrostatic potential, $q_j$ is the charge of species $j$ and $p$ is the momentum of the particles in the initial equilibrium state. Hence, generalized form of three dimensional Kappa- Fermi distribution in the presence of electrostatic potential is

$$f_\kappa(\phi) = \frac{n_{jo}\left\{2-\mu\beta\left(1+\frac{s}{\kappa}\right)\right\}^{\left(\frac{1}{\kappa+s}-\frac{1}{2}\right)} \Gamma\left(\frac{1}{\kappa+s}+1\right)}{\pi^{3/2}\left[\frac{\left(1+\frac{s}{\kappa}\right)\beta}{2m_j}\right]^{-3/2} \Gamma\left(\frac{1}{\kappa+s}-\frac{1}{2}\right)} \left[1+\left\{1+\frac{\beta}{\kappa}\left(\frac{p_j^2}{2m_j}+q_j\phi-\mu\right)\right\}^{\kappa+s}\right]^{-\left(1+\frac{1}{\kappa+s}\right)}.$$

(iii)

For electron species above distribution can be expressed as,

$$f_{\kappa e}(\phi) = \frac{n_{eo}\left\{2 - \mu\beta\left(1 + \frac{s}{\kappa}\right)\right\}^{\left(\frac{1}{\kappa+s} - \frac{1}{2}\right)} \Gamma\left(\frac{1}{\kappa+s} + 1\right)}{\pi^{3/2}\left[\frac{\left(1 + \frac{s}{\kappa}\right)\beta}{2m_e}\right]^{-3/2} \Gamma\left(\frac{1}{\kappa+s} - \frac{1}{2}\right)} \left[1 + \left\{1 + \frac{\beta}{\kappa}\left(\frac{p_e^2}{2m_e} + q_e\phi - \mu\right)\right\}^{\kappa+s}\right]^{-\left(1 + \frac{1}{\kappa+s}\right)}.$$

(iv)

This is the generalized form of three dimensional Kappa-Fermi distribution in the presence of electrostatic potential which can be apply to represent the distribution of Kappa electrons (superthermal).

**Data Availability**

No data was used for the research described in the article.

# Figure captions

Fig.1     (a) Variation of Sagdeev potental $V(\phi)$ with electrostatic potential $\phi$, for different values of Kappa index $\kappa$. The approximate values of $H$ for the three cases are 0.5, 0.7 and 1.09 respectively with $\delta = 1.75$ and $\kappa = 0.5$. (b) Variation of $V(\phi)$ with $\phi$, for different values of number density ratio $\Delta$. The other parameter values are $H = 0.747$, $U = 0.1$, $\delta = 1.75$, $\Delta = 2.0735$ and $\kappa = 0.5$. This plot is corresponding to compressive double layer.

Fig.2     (a) Variation of Sagdeev potental $V(\phi)$ with electrostatic potential $\phi$, for different values of Kappa index $\kappa$. The approximate values of $H$ for the three cases are 0.5, 0.7 and 1.09 respectively with $\delta = 2.88$ and $\kappa = 0.5$. (b) Variation of $V(\phi)$ with $\phi$, for different values of number density ratio $\Delta$ $(\Delta = n_{\kappa e0}/n_{he0})$. The other parameter values are $H = 0.747$, $U = 0.1$, $\delta = 2.88$ $\Delta = 2.0735$ and $\kappa = 0.5$. This plot is corresponding to rarefactive double layer.

Fig.3     (a) Variation of electron acoustic double layer solution $\phi$ with $\eta$, for the different values of $\kappa$ with $U = 0.1$, $\Delta = 2.0735$ and $\delta = 1.75$. (b) Variation of $\phi$ with $\eta$, for the different values of $\Delta$, where the other parameters are $U = 0.1$, $\delta = 1.75$ and $\kappa = 0.5$. This plot is corresponding to compressive double layer.

Fig.4     (a) Variation of electron acoustic double layer solution $\phi$ with $\eta$, for the different values of $\kappa$ with, $\Delta = 2.0735$ and $\delta = 2.88$. (b) Variation of electron acoustic double layer solution $\phi$ with $\eta$, for the different values of $\Delta$, where the other parameters are $\delta = 2.88$ and $\kappa = 0.5$. This plot is corresponding to rarefactive double layer.

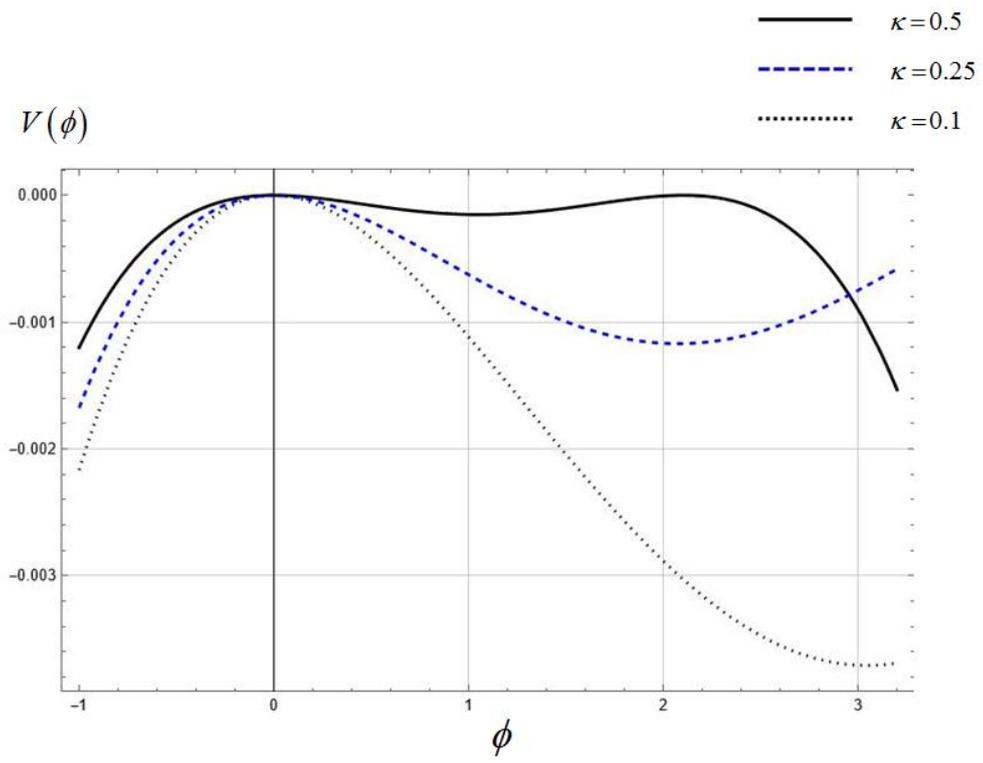

Fig.1(a)

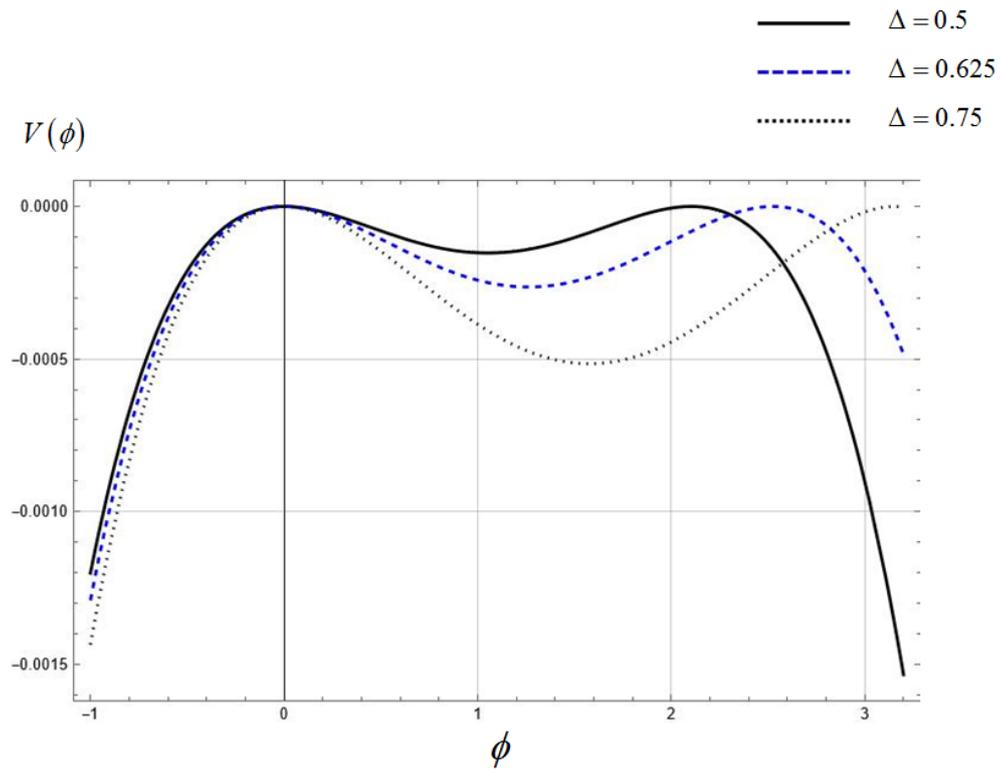

Fig.1(b)

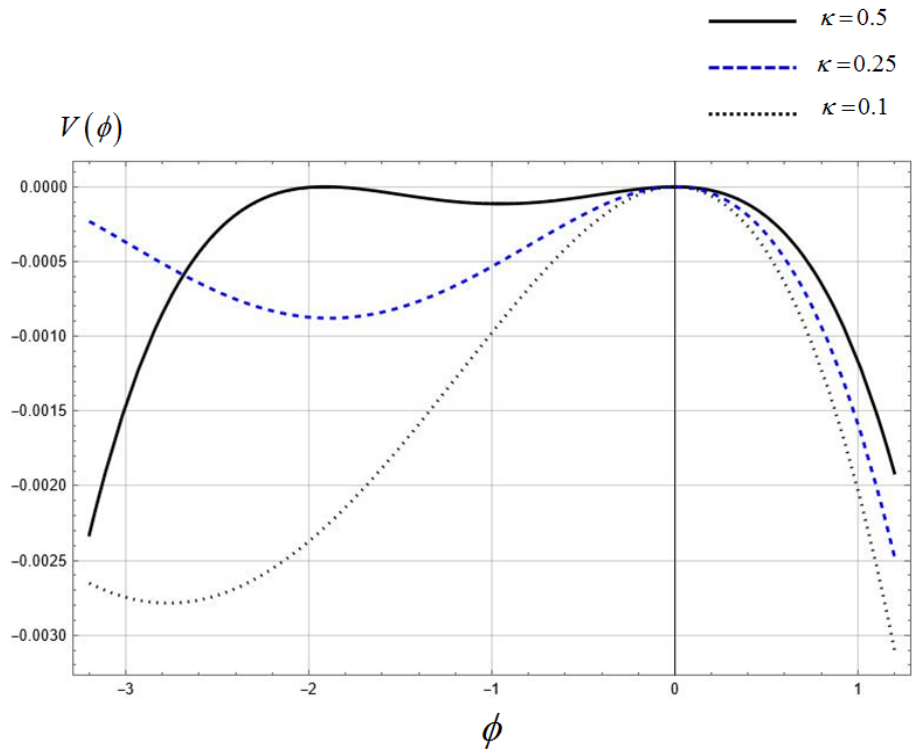

Fig.2(a)

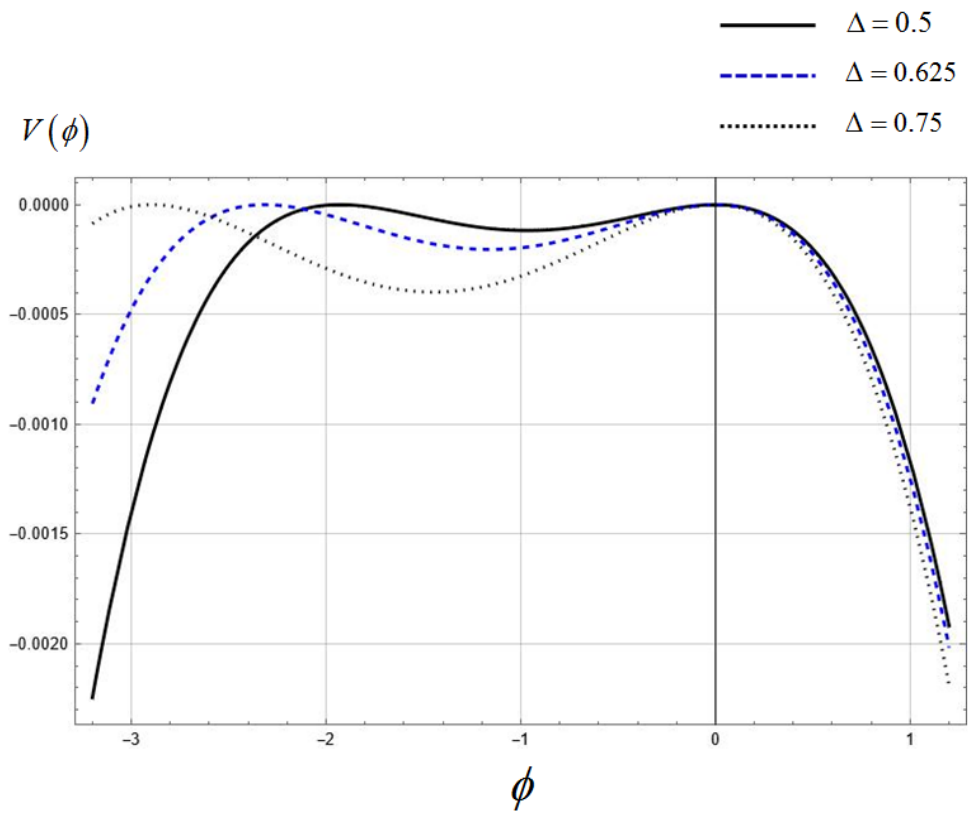

Fig.2(b)

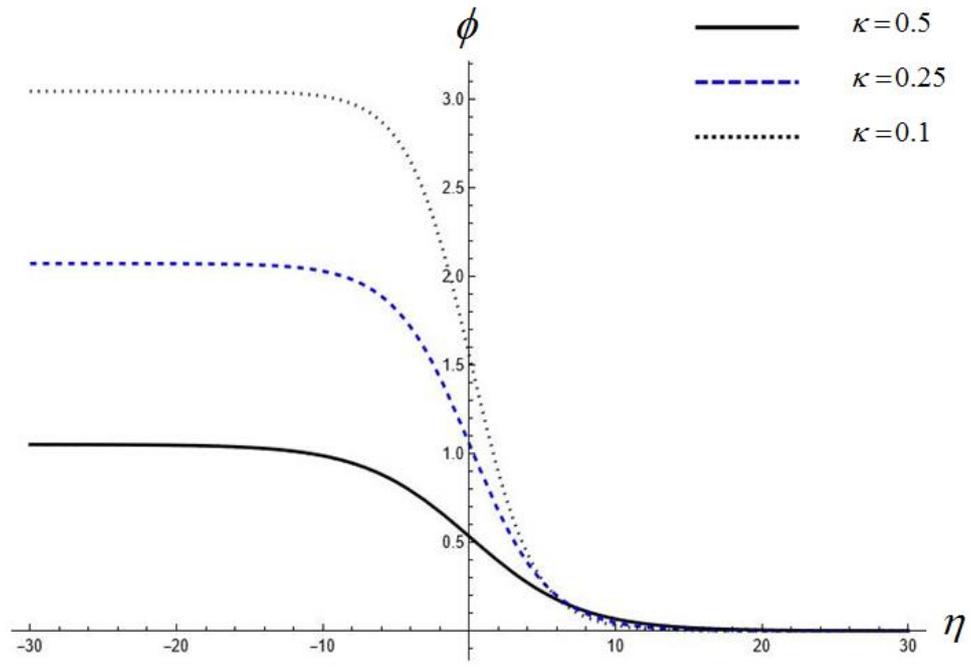

Fig.3(a)

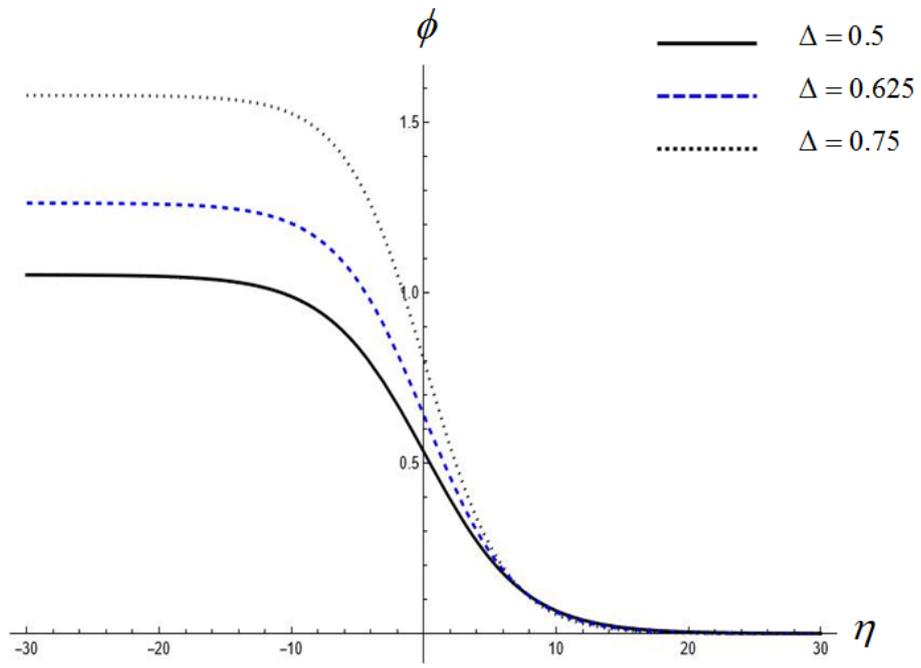

Fig.3(b)

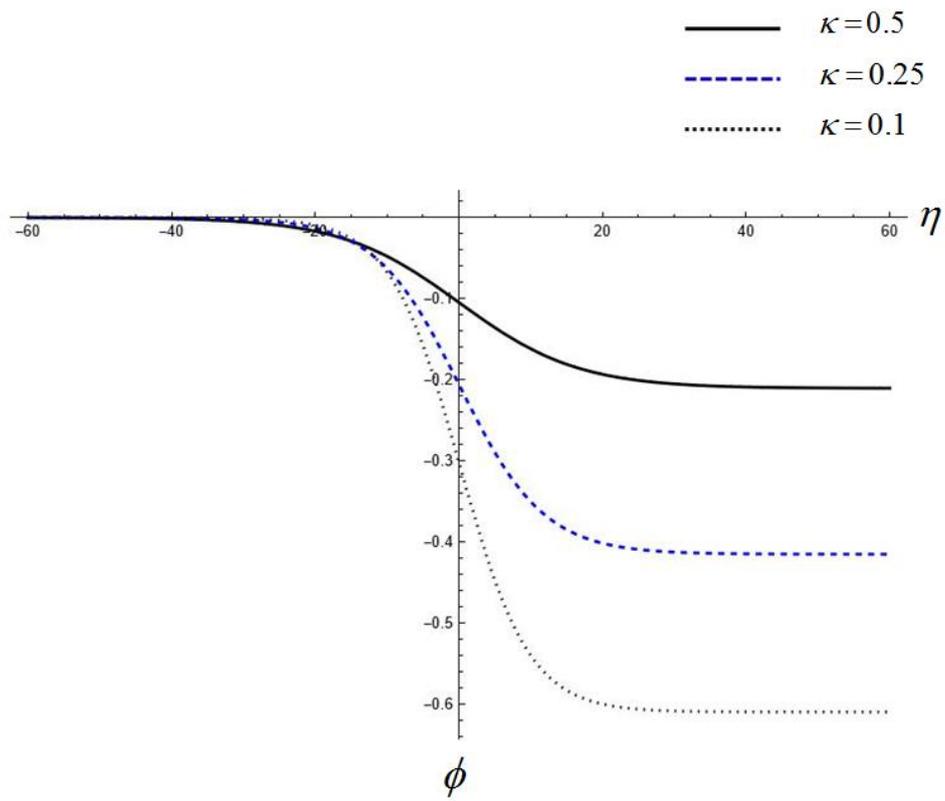

Fig.4(a)

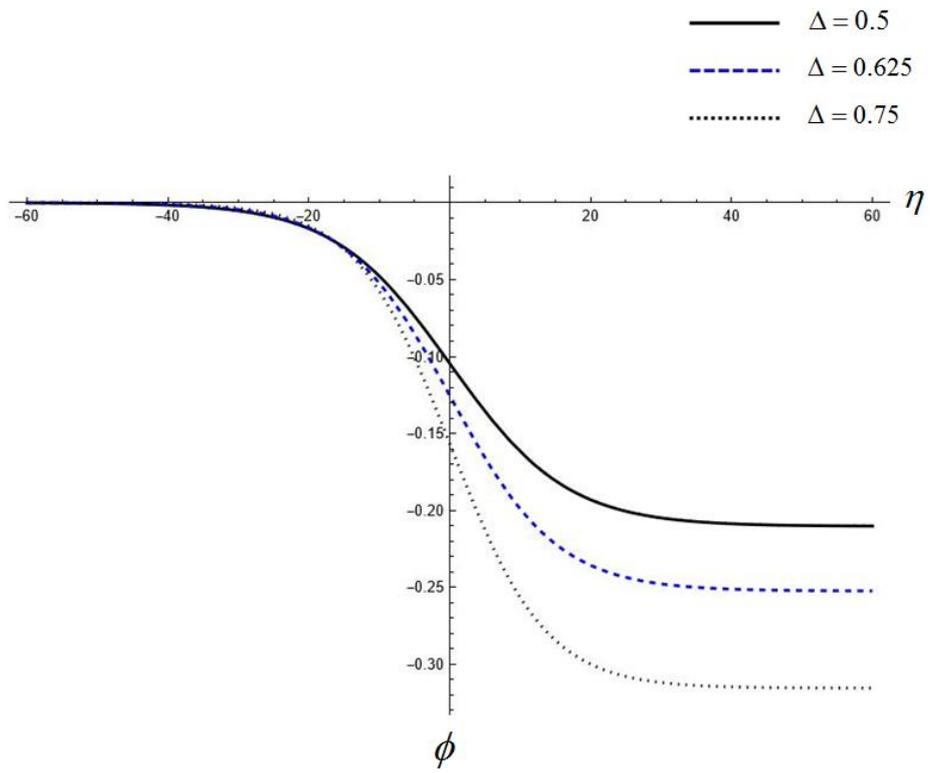

Fig.4(b)